# Free Energy of Nonideal Atomic Plasma

A. L. Khomkin* and I. A. Mulenko**

*IVTAN (Institute of High Temperatures) Scientific Association, Russian Academy of Sciences, Moscow, 125412 Russia
**Ukrainian State Marine Technical University, Nikolaev, Ukraine

Abstract.
An expression is derived for the free energy of nonideal atomic plasma, which corresponds to exact asymptotic expansions. The analytical dependence of correction for the interaction of free charges on the partition function of atom is found. It is demonstrated that the consistent inclusion of the contribution by excited atomic states to the thermodynamic functions of nonideal atomic plasma brings about a significant modification of the conventionally employed equation of state and equation of ionization equilibrium.

## INTRODUCTION

We will treat an atomic hydrogen-like plasma consisting of $N_e$ electrons, $N_i$ ions, and $N_a$ atoms and located in a volume $V$ at a temperature $T$. Assuming that the gas of atoms is ideal, and the free charges weakly interact with one another, the free energy $F$ of the system being treated has the form:

$$\beta F = -N_e \ln\left(\frac{2eV}{N_e \lambda_e^3}\right) - N_i \ln\left(\frac{eV}{N_i \lambda_i^3}\right) - N_a \ln\left(\frac{e\Sigma_a V}{N_a \lambda_a^3}\right) - (N_e + N_i)\Delta F, \quad (1)$$

where $\beta = 1/kT$ is the inverse temperature, $\lambda_k = \left(2\pi\hbar^2 \beta / m_k\right)^{1/2}$ is the thermal wavelength of a particle of the sort $k$ ($k=e,i,a$); $\Sigma_a$ is the internal partition function of atom, and $e$ - is the base of natural logarithm. The last term in Eq. (1) is the correction to the free energy of an ideal-gas mixture, caused by the interaction of free electrons and ions with one another. This correction may be conveniently expressed in terms of the appropriate quantity $\Delta F$ related to temperature and unit charge. Equation (1) describes a system of charges in a fairly wide range of pressures and temperatures, from an ideal gas of atoms to a fully ionized weakly nonideal plasma. This equation serves a basis for the so-called "chemical model" of plasma [1] within which most of concrete calculations by the equations of state and composition of plasma are performed and which allows a generalization to a composite plasma [2]. Note, however, that no unified generally accepted method exists for the notation of free energy for a nonideal atomic plasma [3]. This indeterminacy is due to the fact that different expressions are used in the literature for the finite convergent partition function of atom $S_a$, as well as different expressions for the correction for the interaction of free charges $\Delta F$. Therefore, one can speak of the presence of a double indeterminacy for the free energy of atomic plasma of the form of Eq. (1). This indeterminacy results in numerous versions of the "chemical model" [4] which produce substantially different results in the calculation of equations of state and composition of plasma. Comparison with the results of experiments performed for a plasma of inert gases and vapors of alkali metals [5] only proves the existence of a problem in this case. A surprisingly good agreement with experiment is demonstrated by the ideal-gas approximation ($\Delta F = 0$) even under conditions of a marked Coulomb nonideality [6]. Neither is it possible, within some single version of the "chemical model" (1) to describe the experimental data using both the thermal and caloric equations of state [7].

The same range of states of atomic plasma, i.e., from atomic gas to fully ionized weakly nonideal plasma, allows the description within the "physical model". The main results from the use of this latter model were obtained in [8-10]. In this case, treatment is performed in the grand canonical ensemble by the methods of quantum statistics. For the grand thermodynamic potential $W = -PV$ and total concentration of nuclei $n$, expansions with respect to degrees of activity $z_{e,i} = g_{e,i} \lambda_{e,i}^{-3} e^{\beta \mu_{e,i}}$ and ($g_k$ и $\mu_k$ – denote the statistical weight and chemical potential of particles of the sort $k$, respectively) were obtained up to terms of the order of $z^{5/2}$ и $z^{5/2} \ln z$. Within quadratic terms, these expansions have the following form:

$$-\beta W / V = \beta P = (z_e + z_i)(1 + \frac{\alpha}{3} + \frac{\alpha^2}{8}) + z_e z_i \frac{\lambda_e^3}{2} S_{PL}, \quad (2)$$

$$n = z_e(1 + \frac{\alpha}{2} + \frac{\alpha^2}{4}) + z_e z_i \frac{\lambda_e^3}{2} S_{PL}. \quad (3)$$

Here, $\alpha = \beta e^2 \kappa$, $\kappa = \sqrt{4\pi \beta e^2 (z_e + z_i)}$ – denote the plasma parameter and inverse Debye radius, respectively, calculated in terms of the activities of electrons and ions rather than in terms of the charge concentrations; and $S_{PL}$ – is the Planck--Larkin partition function,

$$S_{PL} = \sum_{n=1}^{\infty} g_n (e^{\beta E_n} - 1 - \beta E_n), \quad (4)$$

where $g_n$, $E_n$ – denote the statistical weight and the binding energy of atomic state with the main quantum number $n$, respectively. Relations (2) and (3) are written with an accuracy of up to terms of the order of $z^2$, and do not contain quantum corrections of the order of $l_e c$. These simplifications, while placing no restrictions on the generality of our inferences, make it possible to appreciably simplify subsequent computations. At high temperatures, equation of state (2), (3) describes the fully ionized plasma ($bP=2n$), and, at lower temperatures, the atomic gas ($bP=n$). The distinguishing feature of relations (2) and (3) is the fact that their use does not require the solution of the problem (that is traditional for the "chemical model") of the choice of the atomic partition function. This is quite natural, because, in deriving Eqs. (2) and (3), we consistently treated the contribution made to thermodynamics by the states of both discrete and continuous spectra of different sign with regard for their screening. Therefore, in contrast to Eq. (1), relations (2) and (3) do not exhibit indeterminacy.

We demonstrate the existence of a class of models in which the free energy of atomic weakly nonideal plasma given by Eq. (1) exactly corresponds to the asymptotic expansions given by Eqs. (2) and (3). This reduces the problem of double indeterminacy of the "chemical model" to the choice of only the atomic partition function. We will follow Volokitin and Kalitkin [11] and conventionally divide the methods of calculation of the atomic partition function used in the literature into three groups, namely, the calculation by formula (4), the application of the nearest neighbor approximation (NNA) (classed with this may be the use of the concepts of critical microfield [12], confluence of lines, etc.), and the calculation of the atomic partition function with regard for all states for which the binding energy of electron exceeds in magnitude the Debye energy of charge in the plasma. In accordance with the inferences of Volokitin and Kalitkin [11] that it is the second approach (i.e., NNA) that best corresponds to the actually observed radiation spectra of atomic plasma, we can derive a unique (in a sense) and quite valid expression for the free energy of nonideal atomic plasma.

1. TRANSITION FROM THE "CHEMICAL" TO "PHYSICAL" MODEL. We will use expression (1) for the free energy of nonideal atomic plasma in the general form and make a transition from the small to grand canonical ensemble. The expression relating the free energy $F$ and the grand thermodynamic potential $W$ has the form:

$$-\frac{bW}{V} = bP = -\frac{bF}{V} + n_e b\mathbf{m}_e + n_i b\mathbf{m}_i + n_a b\mathbf{m}_a. \tag{5}$$

We use relation $\mathbf{m}_k = \partial F/\partial N_k$, to derive, from Eq. (1),

$$b\mathbf{m}_e = -ln\left(\frac{2}{n_e \mathbf{l}_e^3}\right) - n_a \frac{\partial ln\mathbf{S}_a}{\partial n_e} - DF - \frac{G}{2}\frac{\partial DF}{\partial G}, \tag{6}$$

$$\bm{m}_i = -\ln\left(\frac{1}{n_i \bm{l}_i^3}\right) - n_a \frac{\partial \ln S_a}{\partial n_i} - DF - \frac{G}{2}\frac{\partial DF}{\partial G}, \qquad (7)$$

$$\bm{m}_a = -\ln\left(\frac{S_a}{n_a \bm{l}_a^3}\right), \qquad (8)$$

In deriving Eqs. (6) - (8), it was assumed that the correction $DF$ was a function of only the dimensionless plasma parameter $G = \bm{b}e^2 \sqrt{4\bm{p}\bm{b}e^2(n_e + n_i)}$. This is justified by the fact that the Coulomb potential has no characteristic spatial scale.

We will further assume for simplicity that the atomic partition function $S_a$ symmetrically depends on the concentrations of free electrons and ions, because all of the treated thermodynamic relations are characterized by the same dependence. Therefore,

$$\frac{\partial \ln S_a}{\partial n_e} = \frac{\partial \ln S_a}{\partial n_i} = \frac{\partial \ln S_a}{\partial n_{ei}} \qquad (9)$$

We substitute Eqs. (1) and (6) - (9) into (5) to derive

$$-\frac{\bm{b}W}{V} = (n_e + n_i)\left(1 - \frac{G}{2}\frac{\partial DF}{\partial G} - n_a \frac{\partial \ln S_a}{\partial n_{ei}}\right) + n_a = \bm{b}P. \qquad (10)$$

It is now necessary to change in Eq. (10) from concentrations to activities. We use the definition of activity,

$$z_k = \frac{e^{\bm{m}_k} g_k}{\bm{l}_k^3}, \quad k = e, i, \qquad (11)$$

to find, from Eqs. (6) - (8),

$$n_e = z_e \exp\left(DF + \frac{G}{2}\frac{\partial DF}{\partial G} + n_a \frac{\partial \ln S_a}{\partial n_{ei}}\right) \qquad (12)$$

$$n_i = z_i \, exp\left(DF + \frac{G}{2}\frac{\partial DF}{\partial G} + n_a \frac{\partial \ln S_a}{\partial n_{ei}}\right) \qquad (13)$$

$$n_a = z_e z_i \frac{l_e^3}{2} S_a. \qquad (14)$$

In deriving Eq. (14), we further used the equation of ionization equilibrium $m_e + m_i = m_a$, which naturally leads to the Saha formula,

$$n_a = n_e n_i \frac{l_e^3}{2} S_a e^{-bDI}, \qquad (15)$$

where $DI$ is the decrease in the atomic ionization potential in the plasma,

$$bDI = 2DF + G\frac{\partial DF}{\partial G} + 2n_a \frac{\partial \ln S_a}{\partial n_{ei}}. \qquad (16)$$

We substitute Eqs. (12) - (14) into (10) to derive

$$-\frac{bW}{V} = (z_e + z_i)[1 + DF_W(G)] + z_e z_i \frac{l_e^3}{2} S_a, \qquad (17)$$

in which

$$DF_W(G) = \left(1 - \frac{G}{2}\frac{\partial DF}{\partial G} - n_a \frac{\partial \ln S_a}{\partial n_{ei}}\right) e^{bDI/2} - 1. \qquad (18)$$

We equate Eqs. (2) and (17) to derive the equation relating the correction for the interaction of free charges $DF$ to the atomic partition function $S_a$

$$DF_W = \frac{a}{3} + \frac{a^2}{8} - \frac{z_e z_i}{z_e + z_i}\frac{l_e^3}{2}(S_a - S_{PL}). \qquad (19)$$

Equation (19) must be complemented with the formula relating the concentrations and activities or the parameters $a$ and $G$,

$$a^2 = G^2 \exp\left(-DF - \frac{G}{2}\frac{\partial DF}{\partial G} - n_a \frac{\partial \ln S_a}{\partial n_{ei}}\right), \tag{20}$$

which follows from Eqs. (12) and (13).

2. LINEARIZATION OF RESULTANT EQUATIONS. FREE ENERGY OF NONIDEAL ATOMIC PLASMA. Equations (19) and (20) represent a rather complex set of nonlinear dependences. Their direct application is unjustified, because formula (2) is an asymptotic expansion. We linearize Eqs. (17) - (20) to derive expressions relating the correction for the interaction of free charges $DF$ to the atomic partition function $S_a$ to first orders with respect to the plasma parameter $G$.

We linearize Eqs. (18) and (20) to derive

$$DF_W = DF \tag{21}$$

$$a = G. \tag{22}$$

We substitute these expressions into Eq. (19) and derive the sought relation

$$DF = \frac{G}{3} - n_e l_e^3 \frac{1}{4}(S_a - S_{PL}) + O(G^2). \tag{23}$$

Relation (23) eliminates one of the indeterminacies of the "chemical model" and defines the sought class of expressions for the free energy of nonideal atomic plasma, which correspond to exact asymptotic expansions (2) and (3) for some or other choice of atomic partition function.

A number of qualitative inferences follow from Eq. (23). It is only when the Planck - Larkin partition function $S_a = S_{PL}$ is used that the correction to the free energy corresponds to the conventionally employed Debye approximation: $DF = G/3$. If $S_a > S_{PL}$, the correction to the free energy turns out to be less than the Debye correction. If $S_a \gg S_{PL}$, even a change of its sign is possible. We will demonstrate below that all of these options are realized with the selected set of atomic partition functions.

For further computations, we will use the most general expression for the partition function of hydrogen-like atom,

$$S_a = \sum_{n=1}^{\infty} 2n^2 e^{bRy/n^2} w_n, \tag{24}$$

in which $w_n$ – is the cut-off factor providing for the convergence of the partition function and describing some or other mechanism of cutting-off the contribution of the high-lying energy levels of bound electron, and $Ry$ is the energy of the ground state of hydrogen atom. Let $n_s$ be the characteristic value of the main quantum number above which the cutting off is significant. We will use the following characteristic (for example, of NNA) representation for $w_n$:

$$w_n = exp(-\frac{n^6}{n_s^6}). \tag{25}$$

We will identically transform the atomic partition function,

$$S_a = \sum_{n=1}^{\infty} 2n^2 \left( e^{bRy/n^2} - 1 - \frac{bRy}{n^2} \right) w_n + \sum_{n=1}^{\infty} 2n^2 \left( 1 + \frac{bRy}{n^2} \right) w_n. \tag{26}$$

In the first term, we can assume that $w_n=1$, because we treat a weakly nonideal plasma in which the Debye radius and the mean interparticle spacing greatly exceed the Landau length $be^2$. As was observed in [13], the inclusion in this term of the difference between the factor $w_n$ and unity leads to corrections in final expressions of higher order with respect to the plasma parameter, i.e., is an excess of accuracy. As a result, the first term in Eq. (26) corresponds to the Planck - Larkin partition function $S_{PL}$. In the second term, we will similarly change [13] from summation with respect to $n$ to integration, because the main contribution to this term is made by terms with high values of $n$,

$$DS_a = S_a - S_{PL} = 2\int_0^{\infty} n^2 \left( 1 + \frac{bRy}{n^2} \right) exp\left( -(n/n_s)^6 \right) dn. \tag{27}$$

We perform integration and use the notation $G(x)$ for the gamma-function to finally derive

$$DS_a = (bRy)^{3/2} \frac{1}{3} \left[ x_s^{3/2} G(\frac{1}{2}) + x_s^{1/2} G(\frac{1}{6}) \right], \tag{28}$$

$$x_s = \frac{n_s^2}{bRy} = \frac{1}{bE_s}. \tag{29}$$

The parameter $x_s$ in Eq. (28) enables one to investigate the effect of different methods of restriction of the atomic partition function on the free energy and on all of the remaining thermodynamic functions. For the third option of calculation of $S_a$, i.e., allowing for all states of electron with binding energies exceeding the Debye energy or with orbit sizes less than the Debye radius, the choice of $x_s$ is quite obvious: $x_s = 1/G$. In the nearest neighbor approximation, $w_n = exp[-(4p/3)(n_e + n_i)r_n^3]$, where $r_n = a_0 n^2$, $a_0$ is the Bohr radius; therefore, it is $x_s = 2\times 3^{1/3}/G^{2/3}$ that corresponds to this option.

For further analysis, we will introduce the designation for the second term in Eq. (23),

$$D_F = n_e l_e^3 \frac{1}{4}(S_a - S_{PL}). \tag{30}$$

We substitute Eq. (28) into (30) and use the relation $l_e = a_0\sqrt{4pbRy}$, for the correction $DF$ to the free energy associated with the interaction of free charges to derive, in different approximations,

$$DF^{PL} = \frac{G}{3}, \tag{31}$$

$$DF^{NNA} = \frac{G}{3}(1 - \frac{p\sqrt{6}}{16}) - \frac{\sqrt{2p}\,3^{1/6}}{96} G^{5/3} G\left(\frac{1}{6}\right), \tag{32}$$

$$DF^{Deb} = -\frac{p}{96}\sqrt{G} + \frac{G}{3} - \frac{\sqrt{p}}{96} G^{3/2} G\left(\frac{1}{6}\right). \tag{33}$$

When used in expression (1) with the appropriate atomic partition functions, relations (31) - (33) define the sought class of expressions for the free energy of nonideal atomic plasma that correspond to exact asymptotic expansions (2) and (3). The first term of the correction for NNA (32) fully corresponds to that previously obtained in [13]. The Planck--Larkin partition function gives a conventional Debye correction to the free energy according to Eq. (31). However, as was demonstrated in [11], the number of levels realized in this case does not correspond to the number of experimentally observed spectral lines. In addition, the populations of levels, especially of highly excited ones, are not Boltzmann populations. In our opinion, the most preferable option of partition function corresponds to the NNA: in this case, the correction to the free energy remains proportional to the plasma parameter $G$, but with a different numerical coefficient. The physical reasons for this result will be discussed below. The third option of partition function turns out to be completely inadequate, because the correction to the free energy in the $G \to 0$ limit changes sign, which implies the predominance of repulsion in the interaction of free charges.

3. THERMODYNAMIC FUNCTION OF NONIDEAL ATOMIC PLASMA AND ANALYSIS OF THE RESULTS. In order to derive the equation of state and decrease in the ionization potential, one must find an analytical expression for the term in Eqs. (10) and (16) that contains the partition function derivative,

$$D_S = n_a \frac{\partial \ln S_a}{\partial n_{ei}} = n_e n_i \frac{l_e^3}{2} S_a e^{-b\Delta I} \frac{1}{S_a} \frac{\partial DS_a}{\partial x_s} \frac{\partial x_s}{\partial n_{ei}}. \tag{34}$$

We simplify Eq. (34) and introduce the notation $d = \partial \ln x_s / \partial \ln n_{ei}$, derive

$$D_S = n_i \frac{l_e^3}{2} e^{b\Delta I} x_s \frac{\partial DS_a}{\partial x_s} d. \tag{35}$$

Expression (35) is a complex nonlinear equation relative to $D_S$, because, as follows from Eq. (16), the decrease in the ionization potential $b\Delta I$, depends on $D_S$. We resolve this expression by iterations to derive the first terms of series expansion of the sought quantity with respect to the plasma parameter,

$$D_S = G^2 d \frac{\sqrt{p}}{96} \left[ 3 x_s^{3/2} G\left(\frac{1}{2}\right) + x_s^{1/2} G\left(\frac{1}{6}\right) \right]. \tag{36}$$

For the options of the atomic partition function being treated, we have

$$D_S^{PL} = 0 \tag{37}$$

$$D_S^{Deb} = -\frac{\sqrt{p}}{384} \left[ 3G^{1/2} G\left(\frac{1}{2}\right) + G^{3/2} G\left(\frac{1}{6}\right) \right], \quad d = -\frac{1}{4} \tag{38}$$

$$D_S^{NNA} = -\frac{\sqrt{p}}{576} \left[ 6\sqrt{6} GG\left(\frac{1}{2}\right) + 2^{1/3} 6^{1/6} G^{5/3} G\left(\frac{1}{6}\right) \right], \quad d = -\frac{1}{6}. \tag{39}$$

We will restrict ourselves to the NNA option in writing the final relations for the equations of state and for the decrease in the ionization potential,

$$bP = (n_e + n_i)(1 + D_P) + n_a \qquad (40)$$

$$D_P^{NNA} = -\frac{G}{6}\left[\left(1 - \frac{p\sqrt{6}}{8}\right) - G^{2/3}\frac{\sqrt{2p}\,3^{1/6}}{16}G\left(\frac{1}{6}\right)\right] = -\frac{G}{6}\left(0.038 - 1.047 G^{2/3}\right) \qquad (41)$$

$$bDI^{NNA} = G\left[\left(1 - \frac{p\sqrt{6}}{12}\right) - G^{2/3}\frac{\sqrt{2p}\,3^{1/6}}{24}G\left(\frac{1}{6}\right)\right] = G\left(0.359 - 0.698 G^{2/3}\right) \qquad (42)$$

The resultant relations (41) and (42) demonstrate that the consistent inclusion of excited states in determining the free energy of nonideal atomic plasma and the equations of state and ionization equilibrium leads to results which significantly differ from conventional results (see the factor with the plasma parameter). So, the linear term of the correction to the pressure became significantly (by a factor of almost 25) less than the Debye correction. The decrease in the ionization potential behaves similarly. Figures 1--3 illustrate the behavior of corrections to the free energy, pressure, and decrease in the ionization potential depending on $G$ for three options of the "chemical model" treated by us, which differ from one another by the choice of partition function. The NNA option, which apparently must be preferred, demonstrates a significant weakening of the nonideality effects in the case of consistent inclusion of the contribution by excited states. The first terms of expansions in Eqs. (41) and (42) were obtained previously within a different approach [13].

The first Debye correction to the thermodynamic functions of a fully ionized plasma is virial by its nature,

$$bP = 2n\left(1 - \frac{G}{6}\right) \qquad (43)$$

and the range of its validity is quite narrow. It turns out that, rather than being caused by the attraction of free charges, as it was traditionally assumed, the decrease in the ideal-gas pressure described by this correction is due to the formation of highly excited, essentially classical, atoms; naturally, this is a result of attraction as well. In Eq. (43), $n$ is the total concentration of charges. A perfectly analogous situation is observed in the case of atomic gas which allows the formation of bound states, i.e., molecules. In this case, the use of correction calculated in terms of the second virial coefficient is likewise characterized by a very narrow range of validity and describes the pressure decrease due to both the attraction of atoms and the formation of molecules. It is apparent that the extrapolation of virial corrections in systems in which two-particle bound states are formed will never be capable of describing the pressure decrease by a factor of exactly two upon transition of atomic gas into purely molecular one.

For atomic-molecular systems, this problem was solved by Hill [14] who demonstrated that, in order to take into account the interaction of free atoms not bound into molecules, one must extract from the total virial coefficient its part which corresponds to the bound states, with the bound states proper being taken into account in the form of a separate component. The remaining part of the virial coefficient corresponds to the interaction of free atoms.

It proved difficult to directly apply this approach to plasma, because all virial coefficients in Coulomb systems diverge. The derivation of the convergent result given by Eqs. (2) and (3), which actually implies the calculation of the second group coefficient for the plasma, involves the compensation of contributions by the discrete and continuous spectra in a system of like charges. It was as a result of this compensation that the Planck--Larkin partition function appeared in the final expressions, this partition function corresponding to the convergent part of the second group coefficient proportional to the square of concentration. Note that the Planck--Larkin partition function bears no relation to the actual atomic partition function. We have in fact extracted the contribution by highly excited bound states from the virial expansion for a hydrogen plasma.

Consider a hypothetical case of the presence in the plasma of only the excited states $n_a^*$ corresponding to the second term $DS_a$ in Eq. (26). In this case, Eq. (15) of ionization equilibrium will take the form:

$$n_a^* = n_e n_i \frac{l_e^3}{2} DS_a. \tag{44}$$

Assuming that $n_a^* \ll n = n_e + n_a^*$, we derive from the Saha formula (44)

$$n_{e,i} = n(1 - K) \tag{45}$$

$$n_a^* = nK, \tag{46}$$

where

$$K = n \frac{l_e^3}{2} DS_a = \frac{p\sqrt{6}}{24} G + \frac{\sqrt{2p}\, 3^{1/6}}{48} G^{5/3} G\left(\frac{1}{6}\right). \tag{47}$$

We substitute Eqs. (45) and (46) into the expression for pressure (40) and introduce the total concentration of nuclei $n = n_{e,i} + n_a^*$ to derive the traditional expression for pressure (43),

$$bP = 2n(1-K)\left(1 - \frac{G}{6} + \frac{K}{2}\right) + nK \cong 2n\left(1 - \frac{G}{6}\right). \qquad (48)$$

In deriving Eq. (48), we make use of the fact that the correction to pressure (41) turns out to be related to $K$ by

$$D_P^{NNA} = \frac{G}{6} - \frac{K}{2}. \qquad (49)$$

Structure (48) makes it possible to trace the role of excited states in plasma. The formation of bound excited states due to the factor $K$ shows up in three ways. In addition to the apparent decrease in the number of charges (the factor $1 - K$) and the emergence of the second term associated with the contribution by the ideal gas of excited atoms, the Debye correction decreases (the term $K/2$ in the second factor of the first term). The latter occurrence is due to the fact that the calculation of the Debye correction involves a partial inclusion of the part of the phase space of the electron-ion pair that corresponds to negative energies of relative motion, i.e., to excited atomic states.

Therefore, the range of validity of expression (43) may be significantly extended by way of preliminary introduction of the representation of excited atoms even for a formally fully ionized plasma, because excited atoms represent neutral particles. It was the inclusion of this factor that resulted in a decrease in the conventional Debye corrections. The contribution by excited atoms turns out to be of the same order as the contribution by Debye correlations. Therefore, we reach an unexpected conclusion that a plasma exhibiting a marked nonideality is always partly ionized even at high temperatures. The treatment of a fully ionized plasma is possible only in the $G \to 0$ limit. When the Planck - Larkin partition function is used in calculations, the highly excited bound states corresponding to the $DS_a$, contribution relate to the charged component of atomic plasma, i.e., the kinetic and internal energies of these states relate to the charged subsystem. Generally speaking, one can calculate the equation of state the way it is shown above. It is obvious, however, that the thus calculated composition of atomic plasma will be characterized by overestimated concentrations of free charges. In addition, a nonphysical loss of stability will arise in this case in the equations of state and ionization equilibrium due to a significant overestimation of the energy of free charge interaction with plasma. In our opinion, this fact calls for a certain refinement of predictions concerning the instabilities and plasma phase transitions based on the extrapolation of Debye expressions [3].

## CONCLUSION

We have derived a class of expressions for the free energy of nonideal atomic plasma ("chemical model"), which agree with asymptotic expansions. It has been demonstrated that, for the physically most valid option (NNA), the corrections to the thermodynamic functions do not correspond to the Debye approximation. The contribution by highly excited atomic states to the thermodynamics of nonideal atomic plasma was analyzed. Analytical expressions have been derived for corrections to the equations of state and ionization equilibrium, which are characterized by a wider range of validity compared to conventional expressions.

## ACKNOWLEDGMENTS

This study was supported by the L.M. Biberman and V.S. Vorob'ev School (Russian Foundation for Basic Research grant no. 00-15-96529).

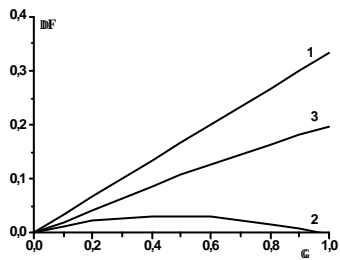 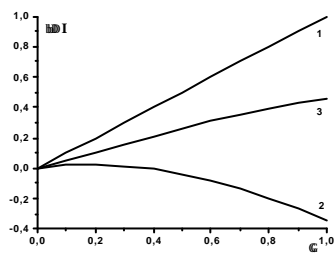 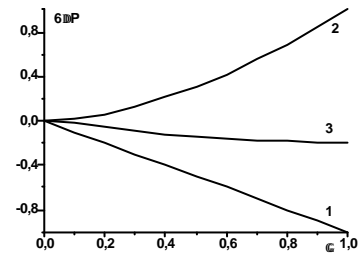

Fig. 1         Fig. 2         Fig. 3

Fig. 1. The correction to the free energy per particle, related to temperature: (*1*) partition function in the Planck - Larkin approximation, (*2*) nearest neighbor approximation, (*3*) summation of all energy levels in the Debye potential.
Fig. 2. The correction to pressure. Designations are the same as in Fig. 1.
Fig. 3. A decrease in the atomic ionization potential. Designations are the same as in Fig. 1.